\documentclass[12pt,a4paper]{article}
\usepackage[utf8]{inputenc}
\usepackage{amsmath}
\usepackage{amsfonts}
\usepackage{amssymb}
\usepackage{graphicx}
\usepackage{tikz}
\usepackage{hyperref}
\usepackage[left=2.00cm, right=2.00cm, top=2.00cm, bottom=2.00cm]{geometry}
\begin{document}
	%\today

\makeatother

\parindent=0cm

\renewcommand{\title}[1]{\vspace{10mm}\noindent{\Large{\bf
			
			#1}}\vspace{8mm}} \newcommand{\authors}[1]{\noindent{\large
		
		#1}\vspace{5mm}} \newcommand{\address}[1]{{ #1\vspace{2mm}}}

\vskip 3mm
\begin{center}
	
\title{New Coalescences for the Painlev\'e Equations}

\date{28-02-2021}	

\authors{
	V.C.C. Alves\footnote{victor.cesar@unesp.br}
}

\address{
	Instituto de F\'{\i}sica Te\'{o}rica-UNESP\\
	Rua Dr Bento Teobaldo Ferraz 271, Bloco II,\\
	01140-070 S\~{a}o Paulo, Brazil}

February 28, 2021
\end{center}
%\begin{center}
%	{\huge {New degeneracies for the Painlev\'e equations}}
%\end{center}
	
\abstract{
The Painlev\'e equations are here connected to other classes of equations with the Painlev\'e Property (Ince's equations) by the same degeneracy procedure that connects the Painlev\'e equations (coalescence). These Ince's equations here are also connected among themselves like in the traditional Painlev\'e's coalescence cascade. Such degeneracy is considered also for the special equations, symmetric equations and B\"acklund transformations.
}

\section{Introduction}
At the end of the 19th century, Painlev\'e and his collaborators worked on the challenge of finding all possible rational second-order ordinary differential equations free of movable critical points (Painlev\'e Property). By a laborious effort on calculations and classification, such work resulted in 50 classes of equations that comprehend all rational second-order equations with the Painlev\'e Property up to M\"obius Transformations. Among these 50, 44 could be linearized or solved by known functions, but 6 could not, therefore they defined new functions, the so-called Painlev\'e Transcendents.

Such a list of the 50 equations can be found in its full presentation on the classical Ince's book of differential equations \cite{ince}.

Since then, Painlev\'e equations have been continuously found in the most diverse areas of mathematical physics and especially for integrable models. Painlev\'e equations appearing as reductions for integrable PDEs is the core of the ARS conjecture \cite{ars}, and are reductions for Dressing Chain of the Schrodinger operator \cite{vaselov}. Together with them, elliptic functions (here, some of Ince's equations) play a major role in the theory of solitons for the traveling wave reduction of integrable models \cite{vaselov}\cite{conte}.

Gambier \cite{gambier24} described that we can reduce the list of 50 equations to only 24 by identifying that the other 26 belong to orbits of these 24 classes of equivalence by birational transformations \cite{conte}.

%\textit{i.e.}, these 24 equations define functions such as Elliptic functions, Painlev\'e transcendents, or trigonometric functions that can be composed to be also the solutions for the remaining 26, therefore they are more fundamental in this sense.

Gambier in his paper \cite{gambier24} lists the 24 equations in a table, such that its correspondence with the usual list of 50 from Ince \cite{ince} is as follows:

\begin{flushleft}
	\begin{tabular}{|c|c|c|c|c|c|c|c|c|c|c|c|c|c}
		\hline 
		Gambier & 1 & 2 & 3 & 4 & 5 & 6 & 7 & 8 & 9 & 10 & 11 & 12 & ...\\ 
		\hline 
		Ince (n) & 1 & 2 & 3 & 4 ($P_I$) & 6 & 5 & 7 & 8 & 9 ($P_{II}$) & 11 & 12 & 13 ($P_{III}$) & ...\\ 
		\hline 
	\end{tabular}
	\vspace{5mm}
	
	\begin{tabular}{c|c|c|c|c|c|c|c|c|c|c|c|c|}
		\hline 
		$\,$...$\,$ & 13 & 14 & 15 & 16 & 17 & 18 & 19 & 20 & 21 & 22 & 23 & 24 \\ 
		\hline 
		$\,\,$...$\,\,$ & 14 & 24 & 27 & 29 & 30 & 31 ($P_{IV}$) & 32 & 37 & 38 & 39 ($P_{V}$) & 49 & 50 ($P_{VI}$) \\ 
		\hline 
	\end{tabular} 
\end{flushleft}

Since Ince is the main reference for them, I will refer to the equations as $I_n$, where the index $n$ is their number on such list. The Painlev\'e equations will be kept as $P_n$ since this is the usual notation.

Composing these 24, we have the 6 Painlev\'e equations ($P_I,...,P_{VI}$), 6 autonomous equations solvable by Elliptic functions ($I_3,\,I_8,\,I_{12},\,I_{30},\,I_{38},\,I_{49}$), 7 absent of parameters ($I_1,\,I_2,\,I_7,\,I_{11},\,I_{29},\,I_{32},\,I_{37}$), and 5 with arbitrary functions ($I_5,\,I_6,\,I_{14},\,I_{24},\,I_{27}$).

P. Painlev\'e himself saw himself that it was possible to connect six Painlev\'e equations noticing that one can transform their variables and parameters artificially introducing a parameter (usually called $\epsilon$) in a so specific way that the limiting procedure of $\epsilon\to 0$ turns an equation into some other. The $P_{VI}$ equation is considered a ``master" equation for the other five since it degenerates into them \cite{plvcoal}.

The Painlev\'e equations are non-linear differential equations with their critical points being just poles, they also have parameters that allow one to construct an infinite chain of solutions for each set of parameters through B\"acklund Transformations \cite{Noumi}. Since the degeneracy cascade mentioned above change the nature of the poles by ``coalescing" them at each step, it is commonly known as coalescence cascade. The coalescence also coalesces the parameters as is seen in section \eqref{symmchap} and in \cite{takano}.

The goal of this paper is to present a full degeneracy cascade connecting not only the usual 6 Painlev\'e equations, which is known but also the other 13 equations. The 5 equations with arbitrary functions, cannot be generated by this limiting procedure like the others, therefore will not be considered. Since the word ``coalescence" brings an idea of quantities coalescing and this may not be the case in some situations here, I will keep it only for the Painlev\'e and autonomous equations, and calling by ``degeneracy", which is more general, the other results presented here.

\section{The traditional coalescence cascade}
The original coalescence cascade for Painlev\'e equations was discovered by Painlev\'e himself \cite{plvcoal} and is here reproduced in order to extend it in the following sections.

As a matter of notation, we will use upper case letters for the equations that degenerate into the other and lower case letters for the resulting equation. The equations will always be composed by a function $w(z)$ with the appropriate parameters $a, b, c, d, e$. $\epsilon$ will be the parameter to be taken to zero.

The cascade will be as follows
\\

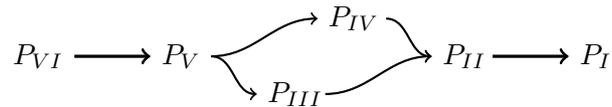
\begin{center}
	\begin{tikzpicture}
\draw[very thick][->]
(0,0) node[left] {$P_{VI}$}
--(1,0) node[right] {$P_V$};
\draw[thick][->]
(1.8,0) to[out=0,in=180] (3.2,0.5) node[right] {$P_{IV}$};
\draw[thick][->]
(1.8,0) to[out=0,in=180] (2.4,-0.5) node[right] {$P_{III}$};
\draw[thick][->]
(4.1,0.5) to[out=0,in=180] (4.7,0) node[right] {$P_{II}$};
\draw[thick][->]
(3.3,-0.5) to[out=0,in=180] (4.7,0);
\draw[very thick][->]
(5.5,0) to[out=0,in=180] (6.5,0) node[right] {$P_{I}$};
\end{tikzpicture}

\end{center}
We start by the $P_{VI}$ equation
\begin{align}\label{p6}
P_{VI}:\,\,W''=\,&\frac{1}{2} \left(\frac{1}{W-Z}+\frac{1}{W}+\frac{1}{W-1}\right) \left(W'\right)^2-\left(\frac{1}{W-Z}+\frac{1}{Z}+\frac{1}{Z-1}\right) W'\\&+\frac{W (W-1) (W-Z)}{Z^2 (Z-1)^2} \left(A +\frac{B  Z}{W^2}+\frac{C  (Z-1)}{(W-1)^2}+\frac{D  (Z-1) Z}{(W-Z)^2}\right)
\end{align}
that with the transformation
\begin{equation}\label{p6p5}
W(Z)=w(z),\qquad Z=1+\epsilon z,\qquad A=a,\qquad B=b,\qquad C=\frac{c}{\epsilon}-\frac{d}{\epsilon^2},\qquad D=\frac{d}{\epsilon}
\end{equation}
followed by $\epsilon\to0$, becomes the $P_V$ equation:
\begin{align}\label{p5}
P_{V}:\,\,w''=\,&\left(\frac{1}{2 w}+\frac{1}{w-1}\right) \left(w'\right)^2-\frac{w'}{z}+\frac{(w-1)^2}{z^2} \left(a  w+\frac{b }{w}\right)+\frac{c  w}{z}+\frac{d  (w+1) w}{w-1}
\end{align}

The $P_V$ equation \eqref{p5} above can degenerate to both $P_{IV}$ and $P_{III}$ with different transformations.

The $P_V$ equation \eqref{p5} with the transformations
\begin{equation}\label{p5p4}
W(Z)=\frac{\epsilon w(z)}{\sqrt{2}},\qquad Z=1+\sqrt{2}\epsilon z,\qquad A=\frac{1}{2\epsilon^4},\qquad B=\frac{b}{4},\qquad C=-\frac{1}{\epsilon^4},\qquad D=\frac{a}{\epsilon^2}-\frac{1}{2\epsilon^4}
\end{equation}
followed by $\epsilon\to0$, becomes the $P_{IV}$ equation:
\begin{align}\label{p4}
P_{IV}:\,\,w''=\,&\frac{3 w^3}{2}+\frac{\left(w'\right)^2}{2 w}+4 w^2 z+2 w \left(z^2-a \right)+\frac{b }{w}
\end{align}

The $P_V$ equation \eqref{p5} with the transformations:
\begin{equation}\label{p5p3}
W(Z)=1+\epsilon z w(z),\qquad Z=z^2,\qquad A=\frac{a}{4\epsilon}+\frac{c}{8\epsilon^2},\qquad B=-\frac{c}{8\epsilon^2},\qquad C=\frac{\epsilon b}{4},\qquad D=\frac{d\epsilon^2}{8}
\end{equation}
followed by $\epsilon\to0$, becomes the $P_{III}$ equation:
\begin{align}\label{p3}
P_{III}:\,\,w''=\,&\frac{\left(w'\right)^2}{w}-\frac{w'}{z}+\frac{a  w^2+b}{z}+c  w^3+\frac{d }{w}
\end{align}

The $P_{IV}$ equation \eqref{p4} with the transformations
\begin{equation}\label{p4p2}
W(Z)=\frac{2^{2/3} w(z)}{\epsilon}+\frac{1}{\epsilon^3},\qquad Z=2^{-2/3}\epsilon z-\frac{1}{\epsilon^3},\qquad A=-2a-\frac{1}{2\epsilon^6},\qquad B=-\frac{1}{2\epsilon^{12}}
\end{equation}
followed by $\epsilon\to0$, becomes the $P_{II}$ equation:
\begin{align}\label{p2}
P_{II}:\,\,w''=\,&2 w^3+w z+a
\end{align}

The $P_{III}$ equation \eqref{p3} with the transformations
\begin{equation}\label{p3p2}
W(Z)=1+2\epsilon w(z),\qquad Z=1+\epsilon^2 z,\qquad A=-\frac{1}{2\epsilon^6},\qquad B=\frac{1}{2\epsilon^6}+\frac{2a}{\epsilon^3},\qquad C=-D=\frac{1}{4\epsilon^6}
\end{equation}
followed by $\epsilon\to0$, becomes the $P_{II}$ equation \eqref{p2}.

The $P_{II}$ equation \eqref{p2} with the transformations
\begin{equation}\label{p2p1}
W(Z)=\epsilon w(z)+\frac{1}{\epsilon^5},\qquad Z=\epsilon^2 z-\frac{6}{\epsilon^{10}},\qquad A=\frac{4}{\epsilon^{15}}
\end{equation}
followed by $\epsilon\to0$, becomes the $P_{I}$ equation:
\begin{align}\label{p1}
P_{I}:\,\,w''=\,&6w^2+z
\end{align}

\section{Autonomous Painlev\'e equations}\label{autopainleve}
These autonomous equations are not solvable by Painlev\'e transcendents, but by Elliptic functions. They are named like that due to their direct connection with each of the Painlev\'e equations both in their form as in degeneracy.

These limits are present explicitly here, which will connect the following equations:
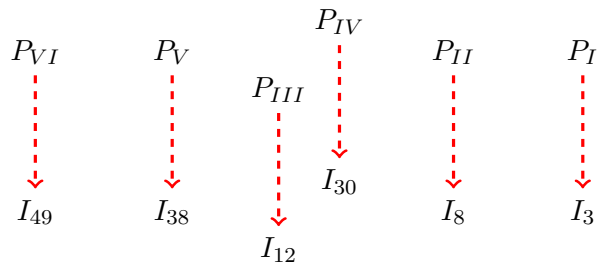
\begin{center}
	\begin{tikzpicture}
\draw [red, dashed, very thick][->]
(-0.4,-0.3) node[black, above] {$P_{VI}$}
-- (-0.4,-1.8) node[black, below] {$I_{49}$};
\draw [red, dashed, very thick][->]
(1.4,-0.3) node[black, above] {$P_{V}$}
-- (1.4,-1.8) node[black, below] {$I_{38}$};
\draw [red, dashed, very thick][->]
(2.8,-0.8) node[black, above] {$P_{III}$}
-- (2.8,-2.3) node[black, below] {$I_{12}$};
\draw [red, dashed, very thick][->]
(3.6,0.1) node[black, above] {$P_{IV}$}
-- (3.6,-1.4) node[black, below] {$I_{30}$};
\draw [red, dashed, very thick][->]
(5.1,-0.3) node[black, above] {$P_{II}$}
-- (5.1,-1.8) node[black, below] {$I_{8}$};
\draw [red, dashed, very thick][->]
(6.8,-0.3) node[black, above] {$P_{I}$}
-- (6.8,-1.8) node[black, below] {$I_{3}$};
\end{tikzpicture}

\end{center}
So, the $P_{VI}$ equation \eqref{p6} under the transformations:
\begin{equation}\label{p6i49}
W(z)=w(z),\quad Z=a+\epsilon z,\quad 
A=\frac{(a-1)^2 a^2 b}{\epsilon^2},\quad 
B=\frac{a c (a-1)^2}{\epsilon^2},\quad 
C=\frac{a^2 d (a-1)}{\epsilon^2},\quad 
D=\frac{ae(a-1) }{\epsilon^2}
\end{equation}
followed by $\epsilon\to0$, becomes the $I_{49}$ equation:
\begin{align}\label{i49}
I_{49}:\,\,w''=\,&\frac{1}{2} \left(\frac{1}{w-a}+\frac{1}{w}+\frac{1}{w-1}\right) \left(w'\right)^2+w (w-1) (w-a) \left(b+\frac{c}{w^2}+\frac{d}{(w-1)^2}+\frac{e }{(w-a)^2}\right)
\end{align}

The $P_{V}$ equation \eqref{p5} under the transformations:
\begin{equation}\label{p5i38}
W(z)=w(z),\quad Z=\epsilon z,\quad 
A=\frac{a}{\epsilon^2},\quad 
B=\frac{b}{\epsilon^2},\quad 
C=\frac{2c-d}{2\epsilon^2},\quad 
D=\frac{d}{2\epsilon^2}
\end{equation}
followed by $\epsilon\to0$, becomes the $I_{38}$ equation:
\begin{align}\label{i38}
I_{38}:\,\,w''=\,&\left(\frac{1}{2 w}+\frac{1}{w-1}\right) \left(w'\right)^2+(w-1) w \left(a (w-1)+\frac{b (w-1)}{w^2}+\frac{c}{w-1}+\frac{d}{(w-1)^2}\right)
\end{align}

The $P_{IV}$ equation \eqref{p4} under the transformations:
\begin{equation}\label{p4i30}
W(z)=\frac{w(z)}{\epsilon},\quad Z=\epsilon z+\frac{a}{\epsilon},\quad 
A=\frac{a^2-b}{\epsilon^2},\quad 
B=\frac{c}{\epsilon^4}
\end{equation}
followed by $\epsilon\to0$, becomes the $I_{30}$ equation:
\begin{align}\label{i30}
I_{30}:\,\,w''=\,&\frac{3 w^3}{2}+\frac{\left(w'\right)^2}{2 w}+4 a w^2+2 b w+\frac{c}{w}
\end{align}

The $P_{III}$ equation \eqref{p3} under the transformations:
\begin{equation}\label{p3i12}
W(z)=\frac{w(z)}{\epsilon},\quad Z=\epsilon z,\quad 
A=\frac{b}{\epsilon},\quad 
B=\frac{c}{\epsilon^3},\quad
C=a,\quad
D=\frac{d}{\epsilon^4}
\end{equation}
followed by $\epsilon\to0$, becomes the $I_{12}$ equation:
\begin{align}\label{i12}
I_{12}:\,\,w''=\,&\frac{\left(w'\right)^2}{w}+a w^3+b w^2+c+\frac{d}{w}
\end{align}

The $P_{II}$ equation \eqref{p2} under the transformations:
\begin{equation}\label{p2i8}
W(z)=\frac{w(z)}{\epsilon},\qquad Z=\epsilon z+\frac{a}{\epsilon^2},\qquad 
A=\frac{b}{\epsilon^3}
\end{equation}
followed by $\epsilon\to0$, becomes the $I_{8}$ equation:
\begin{align}\label{i8}
I_{8}:\,\,w''=\,&2 w^3+a w+b
\end{align}

The $P_{I}$ equation \eqref{p1} under the transformations:
\begin{equation}\label{p1i3}
W(z)=\frac{w(z)}{\epsilon^2},\qquad Z=\epsilon z+\frac{1}{2\epsilon^4}
\end{equation}
followed by $\epsilon\to0$, becomes the $I_{3}$ equation:
\begin{align}\label{i3}
I_{3}:\,\,w''=\,&6w^2+\frac{1}{2}
\end{align}

\section{Coalescence cascade between the autonomous Painlev\'e}
Now continuing the same reasoning as Painlev\'e, we are also able to connect the autonomous equations in a cascade in the same fashion as the original Painlev\'e coalescence cascade.

In fact, the transformations for the variables $w$ and $z$ are very similar in both cases.

The cascade will follow the same lines as the Painlev\'e ones:
\begin{center}
	
\begin{tikzpicture}
\draw[very thick][->]
(0,0) node[left] {$I_{49}$}
--(1,0) node[right] {$I_{38}$};
\draw[thick][->]
(1.8,0) to[out=0,in=180] (3.2,0.5) node[right] {$I_{30}$};
\draw[thick][->]
(1.8,0) to[out=0,in=180] (2.4,-0.5) node[right] {$I_{12}$};
\draw[thick][->]
(4.1,0.5) to[out=0,in=180] (4.7,0) node[right] {$I_{8}$};
\draw[thick][->]
(3.3,-0.5) to[out=0,in=180] (4.7,0);
\draw[very thick][->]
(5.5,0) to[out=0,in=180] (6.5,0) node[right] {$I_{3}$};
\end{tikzpicture}
\end{center}

So, we start by the $I_{49}$ equation \eqref{i49}, that under the transformations:
%\begin{equation}
%\begin{split}
\begin{gather}\label{i49i38}
W(Z)=\frac{w(z)}{w(z)-1},\qquad Z=\epsilon z,\nonumber\\
A=\frac{a}{\epsilon},\qquad
B=\frac{4 \epsilon  (a-b \epsilon +b)+d (\epsilon -1)}{\epsilon ^2 \left(\epsilon ^2-3 \epsilon +2\right)},\quad
C=\frac{b}{\epsilon},\qquad
D=\frac{a}{(1-\epsilon)^2\epsilon}\\
E=\frac{a \epsilon  \left(\epsilon ^3-4 \epsilon ^2+4\right)-\epsilon  (\epsilon -1)^2 \left(b \left(\epsilon ^3-8 \epsilon ^2+12 \epsilon -4\right)-c (\epsilon -2)\right)-d (\epsilon -1)^4}{(\epsilon -2) (\epsilon -1)^2 \epsilon ^4}\nonumber
\end{gather}
%\end{split}
%\end{equation}
followed by $\epsilon\to0$, becomes the $I_{38}$ equation \eqref{i38}.

The $I_{38}$ equation \eqref{i38} under the transformations:
\begin{gather}\label{i38i30}
W(Z)=\frac{\epsilon w(z)}{\sqrt{2}},\quad Z=\epsilon z,\quad
A=\frac{1}{\epsilon^4},\quad
B=\frac{C}{2},\nonumber\\
C=-\frac{8 \sqrt{2} a \epsilon +c \epsilon ^4+6}{2 \epsilon ^4},\quad
D=\frac{-4 \sqrt{2} A \epsilon -2 B \epsilon ^2+G \epsilon ^4-2}{\epsilon ^4}
\end{gather}
followed by $\epsilon\to0$, becomes the $I_{30}$ equation \eqref{i30}.

The $I_{38}$ equation \eqref{i38} under the transformations:
\begin{equation}\label{i38i12}
W(Z)=1+\epsilon w(z),\quad Z=z,\quad
A=\frac{a}{2\epsilon^2},\quad
B=\frac{b}{2}-\frac{a}{2\epsilon^2}+\epsilon(2d\epsilon-c),\quad
C=\epsilon(2d\epsilon-c),\quad
D=d\epsilon^2
\end{equation}
followed by $\epsilon\to0$, becomes the $I_{12}$ equation \eqref{i12}.

The $I_{30}$ equation \eqref{i30} under the transformations:
\begin{equation}\label{i30i8}
W(Z)=\frac{w(z)}{\epsilon}+\frac{1}{\epsilon^3},\quad Z=\epsilon z-\frac{1}{\epsilon^3},\quad
A=-\frac{1}{\epsilon^3}-\frac{a \epsilon}{12},\quad
B=\frac{3+a\epsilon^4}{2\epsilon^6},\quad
C=-\frac{4 a \epsilon ^4-6 b \epsilon ^6+3}{6 \epsilon ^{12}}
\end{equation}
followed by $\epsilon\to0$, becomes the $I_{8}$ equation \eqref{i8}.

The $I_{12}$ equation \eqref{i12} under the transformations:
\begin{equation}\label{i12i8}
W(Z)=1+2\epsilon w(z),\quad Z=1+\epsilon^2 z,\quad
A=\frac{1}{4\epsilon^6},\quad
B=-\frac{1}{2\epsilon^6},\quad
C=\frac{2 a \epsilon ^2+1}{2 \epsilon ^6},\quad
D=\frac{-4 a \epsilon ^2+8 b \epsilon ^3-1}{4 \epsilon ^6}
\end{equation}
followed by $\epsilon\to0$, becomes the $I_{8}$ equation \eqref{i8}.

The $I_{8}$ equation \eqref{i8} under the transformations:
\begin{equation}\label{i8i3}
W(Z)=\epsilon w(z)+\frac{1}{\epsilon^5},\quad Z=\epsilon^2 z-\frac{6}{\epsilon^{10}},\quad
A=-\frac{6}{\epsilon^{10}},\quad
B=\frac{8+\epsilon^{12}}{2\epsilon^{15}}
\end{equation}
followed by $\epsilon\to0$, becomes the $I_{3}$ equation \eqref{i3}.

\section{Degeneracy to parameterless equations}
The parameterless equations have a very simple form, such that they in some cases are just the corresponding Painlev\'e equation with all parameters set to zero. For all the cases the transformation for both Painlev\'e and corresponding autonomous Painlev\'e is  the same, with the only exception being $P_{VI}$ and $I_{49}$ to $I_{32}$. Also, if some parameter is not specified, it will be naturally deleted by the transformation.

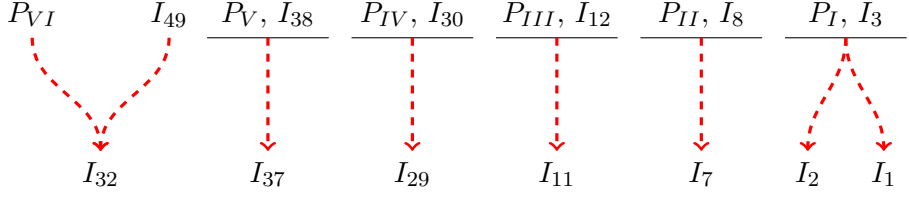
\begin{center}
	
%\begin{tikzpicture}
%\draw [red, dashed, very thick][->]
%(0.0,0.0) node[black, above] {$P_{VI}$}
%to[out=270,in=90] (0.9,-1.5) node[black, below] {$I_{32}$};
%\draw [red, dashed, very thick][->]
%(1.8,0.0) node[black, above] {$I_{49}$}
%to[out=270,in=90] (0.9,-1.5);
%\draw [thin]
%(3.3,0.0) -- (4.9,0.0) node[black,pos=0.5, above] {$P_V,\,I_{38}$};
%\draw [red, dashed, very thick][->]
%(4.1,0.0) -- (4.1,-1.5) node[black, below] {$I_{37}$};
%\draw [thin]
%(6.2,0.0) -- (7.8,0.0) node[black,pos=0.5, above] {$P_{IV},\,I_{30}$};
%\draw [red, dashed, very thick][->]
%(7.0,0.0) -- (7.0,-1.5) node[black, below] {$I_{29}$};
%\draw [thin]
%(9.1,0.0) -- (10.7,0.0) node[black,pos=0.5, above] {$P_{III},\,I_{12}$};
%\draw [red, dashed, very thick][->]
%(9.9,0.0) -- (9.9,-1.5) node[black, below] {$I_{11}$};
%\draw [thin]
%(12.0,0.0) -- (13.6,0.0) node[black,pos=0.5, above] {$P_{II},\,I_{8}$};
%\draw [red, dashed, very thick][->]
%(12.8,0.0) -- (12.8,-1.5) node[black, below] {$I_{7}$};
%\draw [thin]
%(14.9,0.0) -- (16.5,0.0) node[black,pos=0.5, above] {$P_I,\,I_{3}$};
%\draw [red, dashed, very thick][->]
%(15.7,0.0) to[out=270,in=90] (15.2,-1.5) node[black, below] {$I_{2}$};
%\draw [red, dashed, very thick][->]
%(15.7,0.0) to[out=270,in=90] (16.2,-1.5) node[black, below] {$I_{1}$};
%\end{tikzpicture}

\begin{tikzpicture}
\draw [red, dashed, very thick][->]
(0.0,0.0) node[black, above] {$P_{VI}$}
to[out=270,in=90] (0.9,-1.5) node[black, below] {$I_{32}$};
\draw [red, dashed, very thick][->]
(1.8,0.0) node[black, above] {$I_{49}$}
to[out=270,in=90] (0.9,-1.5);
\draw [thin]
(2.3,0.0) -- (3.9,0.0) node[black,pos=0.5, above] {$P_V,\,I_{38}$};
\draw [red, dashed, very thick][->]
(3.1,0.0) -- (3.1,-1.5) node[black, below] {$I_{37}$};
\draw [thin]
(4.2,0.0) -- (5.8,0.0) node[black,pos=0.5, above] {$P_{IV},\,I_{30}$};
\draw [red, dashed, very thick][->]
(5.0,0.0) -- (5.0,-1.5) node[black, below] {$I_{29}$};
\draw [thin]
(6.1,0.0) -- (7.7,0.0) node[black,pos=0.5, above] {$P_{III},\,I_{12}$};
\draw [red, dashed, very thick][->]
(6.9,0.0) -- (6.9,-1.5) node[black, below] {$I_{11}$};
\draw [thin]
(8.0,0.0) -- (9.6,0.0) node[black,pos=0.5, above] {$P_{II},\,I_{8}$};
\draw [red, dashed, very thick][->]
(8.8,0.0) -- (8.8,-1.5) node[black, below] {$I_{7}$};
\draw [thin]
(9.9,0.0) -- (11.5,0.0) node[black,pos=0.5, above] {$P_I,\,I_{3}$};
\draw [red, dashed, very thick][->]
(10.7,0.0) to[out=270,in=90] (10.2,-1.5) node[black, below] {$I_{2}$};
\draw [red, dashed, very thick][->]
(10.7,0.0) to[out=270,in=90] (11.2,-1.5) node[black, below] {$I_{1}$};
\end{tikzpicture}

\end{center}
The $P_{VI}$ equation \eqref{p6} under the transformations
\begin{equation}\label{p6i32}
W(Z)=\epsilon w(z),\qquad Z=\frac{e^z}{\epsilon},\qquad B=-\frac{\epsilon^2}{2},\qquad C=\epsilon
\end{equation}
(here $e^z$ is the exponential of $z$) followed by $\epsilon\to0$ becomes the $I_{32}$ equation:
\begin{equation}\label{i32}
I_{32}:\qquad w''=\frac{w'^2-1}{2 w}
\end{equation}

The $I_{49}$ equation \eqref{i49} under the transformations
\begin{equation}\label{i49i32}
W(Z)=\epsilon w(z),\qquad Z=\frac{z}{\epsilon},\qquad A=a,\qquad B=\frac{\epsilon^2}{a},\qquad C=-\frac{\epsilon^4}{2a},\qquad
D=\frac{\epsilon^2}{a},\qquad E=-2a\epsilon^2
\end{equation}
followed by $\epsilon\to0$ becomes the $I_{32}$ equation \eqref{i32}.

The $P_{V}$ \eqref{p5} and $I_{38}$ \eqref{i38} under the transformations
\begin{equation}\label{p5i37}
W(Z)=\epsilon w(z),\qquad Z=\epsilon\ln(z)
\end{equation}
followed by $\epsilon\to0$ becomes the $I_{37}$ equation:
\begin{equation}\label{i37}
I_{37}:\qquad w''=\left(\frac{1}{2 w}+\frac{1}{w-1}\right) w'^2-\frac{w'}{z}
\end{equation}

The $P_{IV}$ \eqref{p4} and $I_{30}$ \eqref{i30} under the transformations
\begin{equation}\label{p4i29}
W(Z)=\frac{w(z)}{\epsilon},\qquad Z=\epsilon z
\end{equation}
followed by $\epsilon\to0$ becomes the $I_{29}$ equation:
\begin{equation}\label{i29}
I_{29}:\qquad w''=\frac{w'^2}{2 w}+\frac{3 w^3}{2}
\end{equation}

The $P_{III}$ \eqref{p3} and $I_{12}$ \eqref{i12} under the transformations
\begin{equation}\label{p3i11}
W(Z)=w(z),\qquad Z=\epsilon z
\end{equation}
followed by $\epsilon\to0$ becomes the $I_{11}$ equation:
\begin{equation}\label{i11}
I_{11}:\qquad w''=\frac{w'^2}{w}
\end{equation}

The $P_{II}$ \eqref{p2} and $I_{8}$ \eqref{i8} under the transformations
\begin{equation}\label{p2i7}
W(Z)=\frac{w(z)}{\epsilon},\qquad Z=\epsilon z
\end{equation}
followed by $\epsilon\to0$ becomes the $I_{7}$ equation:
\begin{equation}\label{i7}
I_7:\qquad w''=2w^3
\end{equation}

The $P_{I}$ \eqref{p1} and $I_{3}$ \eqref{i3} under the transformations
\begin{equation}\label{p1i2}
W(Z)=\frac{w(z)}{\epsilon^2},\qquad Z=\epsilon z
\end{equation}
followed by $\epsilon\to0$ becomes the $I_{2}$ equation:
\begin{equation}\label{i2}
I_2:\qquad w''=6w^2
\end{equation}

The $P_{I}$ \eqref{p1} and $I_{3}$ \eqref{i3} under the transformations
\begin{equation}\label{p1i1}
W(Z)=\epsilon w(z),\qquad Z=\epsilon z
\end{equation}
followed by $\epsilon\to0$ becomes the $I_{1}$ equation:
\begin{equation}\label{i1}
I_1:\qquad w''=0
\end{equation}

\section{Riccati equations}
Painlev\'e equations are known to have Riccati equations as special solutions, such being classical special functions.

These special functions also have a coalescence relation among them and it was fully exposed at \cite{tamiz}, so in this discussion I will only focus on the coalescence limits leading to the corresponding autonomous equations of these special functions, using exactly the same limits as in section \eqref{autopainleve}.

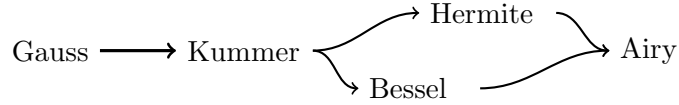
\begin{center}
		\begin{tikzpicture}
\draw[very thick][->]
(0,0) node[left] {Gauss}
--(1,0) node[right] {Kummer};
\draw[thick][->]
(2.8,0) to[out=0,in=180] (4.2,0.5) node[right] {Hermite};
\draw[thick][->]
(2.8,0) to[out=0,in=180] (3.4,-0.5) node[right] {Bessel};
\draw[thick][->]
(6.0,0.5) to[out=0,in=180] (6.7,0) node[right] {Airy};
\draw[thick][->]
(5.0,-0.5) to[out=0,in=180] (6.7,0);
\end{tikzpicture}

\end{center}
I also present these special cases of Painlev\'e equations in a different form, showing that in all cases we are able to naturally obtain the conditions for the Riccati equations.

Here I also follow the generalization used by \cite{smith}, where the parameters $\theta_0, \theta_1, \theta_2$ are signal choices, \textit{i.e.}, they are $\pm 1$.

We begin by the most complex case, the $P_{VI}$ equation \eqref{p6}, which with the special choice of parameters:
\begin{equation}\label{p6spar}
a = \frac{\alpha^2}{2},\quad b = -\frac{1}{2} (\beta-\theta_0 (\gamma \theta_1 \theta_2+1)){}^2,\quad c = \frac{1}{2} (\alpha \theta_1+\gamma){}^2,\quad d = \frac{1}{2} \left(1-\beta^2\right)
\end{equation}
can be written as:
\begin{equation}
(P_{VI}):\quad F'(z)=F(z)^2 \left(\frac{z}{2 w(z)}-\frac{w(z)}{2}\right)+F(z) \left(-\frac{\theta _0 \left(-\beta +\gamma \theta _0 \theta _1 \theta _2+\theta _0\right)}{(z-1) w(z)}-\frac{\alpha  \theta _2 w(z)+2 z-1}{(z-1) z}\right)
\end{equation}
where
\begin{align}
F(z)=&\frac{1}{(w(z)-1) (w(z)-z)}\left(w'(z)-\left(\frac{\alpha  \theta _2 w(z)^2}{(z-1) z}+\frac{\theta _0 \left(\beta -\theta _0 \left(\gamma  \theta _1 \theta _2+1\right)\right)}{z-1}\right.\right.\nonumber\\
&\left.\left.+\frac{-\alpha  \gamma  \theta _1+\beta  \gamma  \theta _0 \theta _1 \theta _2-\gamma ^2+z w(z) \left((\alpha -\beta ) (\alpha +\beta )+\alpha  \gamma  \theta _1-\alpha  \theta _2+\beta  \gamma  \theta _0 \theta _1 \theta _2+\beta  \theta _0-\gamma  \theta _1 \theta _2\right)}{(z-1) z \left(-\alpha  \theta _2+\beta  \theta _0-\gamma  \theta _1 \theta _2\right)}\right)\right)
\end{align}

Notice here for this notation that the $F(z)=0$ is a also a Riccati equation, and in fact that is the one treated extensively in literature as the special case for $P_{VI}$.
%We also observe that the $P_{VI}$ written in this form is a ``Riccati of a Riccati".

$F(z)=0$ can be linearized with a Cole-Hopf transformation:
\[w(z)=-\frac{z(z-1)}{\alpha\theta_2}\frac{u'(z)}{u(z)}\]
having its solutions given in terms of Gauss-Hypergeometric functions.

The coalescence reduction of $F(z)=0$ with the same limits as of $P_{VI}\to I_{49}$ \eqref{p6i49} yields:
\begin{equation}
w'(z)=\sqrt{2b} \theta_2 w(z)^2+\theta_0 \sqrt{-2a c}-\frac{w(z) (a (a b+b+c+d)+c+e)}{\sqrt{2} \left(a \sqrt{b} \theta_2-\theta_0 \sqrt{-a c}\right)}
\end{equation}

For completeness of notation, I will express here the others Painlev\'e equations in the same form as above, but this time I will just make some notation changes, as for $a$ and $d$ in \eqref{p6spar} and will NOT impose conditions on them \textit{a priori}, like those upon $b$ and $c$ in \eqref{p6spar}.
\\

The equation $P_V$ \eqref{p5} can be rewritten without lose of generality with parameters
\[a = \frac{\alpha ^2}{2},\qquad b = -\frac{\beta ^2}{2},\qquad\delta = -\frac{1}{2}\] by
\begin{align}\label{p5sp}
(P_{V}):\quad F'(z)=&\,\,F(z)^2 (1-w(z))+F(z) \left(-\frac{\alpha  \theta _2 w(z)}{z}+\frac{\theta _1}{w(z)-1}+\frac{\alpha  \theta _2-1}{z}\right)+\nonumber\\
&\,\,\frac{\alpha  \theta _1 \theta _2-\beta  \theta _0 \theta _1+c-\theta _1}{2 z (w(z)-1)}
\end{align}
with
\begin{equation}\label{p5f}
F(z)=\frac{1}{2 (w(z)-1) w(z)}\left(w'(z)-\left(w(z) \left(\theta _1-\frac{\alpha  \theta _2+\beta  \theta _0}{z}\right)+\frac{\alpha  \theta _2 w(z)^2}{z}+\frac{\beta  \theta _0}{z}\right)\right)
\end{equation}

Now if we set $F(z)=0$, \eqref{p5f} is solvable by Confluent Hypergeometric functions (or Whittaker functions) and \eqref{p5sp} imposes the condition on the parameters:
\[c= \theta _1 \left(-\alpha  \theta _2+\beta  \theta _0+1\right)\]

We can see here that expressing $P_{V}$ as \eqref{p5sp} give us automatically the conditions on the parameters required for it to be a special function.

The coalescence reduction of $F(z)=0$ with the same limits as of $P_{V}\to I_{38}$ \eqref{p5i38} yields:
\begin{equation}
w'= a \theta _2 w^2-w \left(a \theta _2+b \theta _0\right)+b \theta _0
\end{equation}
\\

The equation $P_{IV}$ \eqref{p4} can be rewritten without lose of generality with parameter
\[b = -\frac{\beta ^2}{2}\] by
\begin{equation}\label{p4sp}
(P_{IV}):\quad F'(z)=-F(z)^2-2 F(z) \left(\theta_1 w(z)+\theta_1 z\right)+\frac{1}{2} \left(-2 a -\beta  \theta _0 \theta_1-2 \theta_1\right)
\end{equation}
with
\begin{equation}\label{p4f}
F(z)=\frac{1}{2 w(z)}\left(w'(z)-\beta  \theta _0-\theta_1 w(z)^2-2 \theta_1 z w(z)\right)
\end{equation}

Now if we set $F(z)=0$, \eqref{p4f} is solvable by Hermite functions and \eqref{p4sp} imposes the condition on the parameter:
\[a=\frac{1}{2} \left(-\beta  \theta _0 \theta_1-2 \theta_1\right)\]

The coalescence reduction of $F(z)=0$ with the same limits as of $P_{IV}\to I_{30}$ \eqref{p4i30} yields:
\begin{equation}
w'=\theta_1 w^2+2 a\theta_1 w+\sqrt{-2c} \theta_0
\end{equation}
\\

The equation $P_{III}$ \eqref{p3} can be rewritten without lose of generality with parameter
\[c=1,\qquad d=-1\] by
\begin{equation}\label{p3sp}
(P_{III}):\quad F'(z)=F(z) \left(\frac{\theta_0}{w(z)}-\theta _1 w(z)-\frac{1}{z}\right)+\frac{a \theta _1 \theta_0+b-2 \theta_0}{z w(z)}
\end{equation}
with
\begin{equation}\label{p3f}
F(z)=\frac{1}{ w(z)}\left(w'(z)-\frac{\left(a \theta _1-1\right) w(z)}{z}-\theta_0-\theta _1 w(z)^2\right)
\end{equation}

Now if we set $F(z)=0$, \eqref{p3f} is solvable by Bessel functions and \eqref{p3sp} imposes the condition on the parameter:
\[b= 2 \theta_0-a \theta _1 \theta_0\]

The coalescence reduction of $F(z)=0$ with the same limits as of $P_{III}\to I_{12}$ \eqref{p3i12} yields:
\begin{equation}
w'= \theta _1 w^2+b \theta _1 w
\end{equation}
\\

The equation $P_{II}$ \eqref{p2} can be rewritten without lose of generality by
\begin{equation}\label{p2sp}
(P_{II}):\quad \theta_0 F'(z)=-2w(z)F(z)+a+\frac{\theta_0}{2}
\end{equation}
with
\begin{equation}\label{p2f}
F(z)=\theta_0  w'(z)-w(z)^2-\frac{z}{2}
\end{equation}

Now if we set $F(z)=0$, \eqref{p2f} is solvable by Airy functions and \eqref{p2sp} imposes the condition on the parameter:
\[a=-\frac{\theta_0}{2}\]

The coalescence reduction of $F(z)=0$ with the same limits as of $P_{II}\to I_{8}$ \eqref{p2i8} yields:
\begin{equation}
w'= \theta _0 w^2+\frac{a \theta _0}{2}
\end{equation}

One can immediately observe that all of these autonomous limits of special functions have the general form
\[w'=c_2w^2+c_1w+c_0\]
and also that the number of parameters decrease from higher equations to lower ones, therefore the degeneracy cascade here is trivial, only needing sometimes to do a shift b y a constant like $w\to w+C$, and then match the parameters.

The linearization of such general form via Cole-Hopf transformation yields:
\begin{equation}\label{autospecial}
u''=K_1u'+K_2u \Longrightarrow u(z)=c_1 e^{\frac{1}{2} z \left(K_1-\sqrt{K_1^2+4 K_2}\right)}+c_2 e^{\frac{1}{2} z \left(K_1+\sqrt{K_1^2+4 K_2}\right)}.
\end{equation}

The form \eqref{autospecial} was expected, since all the autonomous Painlev\'e have solutions in terms of Jacobi Elliptic Functions and \eqref{autospecial} is the corresponding Riccati form for them.

\section{Coalescence for the symmetric $P_{IV}$ and $P_V$}\label{symmchap}
In this section I present some of the previous results for $P_{IV}$ and $P_V$ but in the framework of their symmetric equations.

I show how the traditional coalescence appears as additional terms in the equations of motion and how the coalescence to autonomous Painlev\'e appears as a constraint ($\sigma=0$) on the parameters.

The idea behind it was first noticed by \cite{vaselov}, in the context of dressing chains, however the link with Ince's equations, and coalescence was not considered.

\subsection{$P_{IV}$}
\subsubsection{Traditional symmetric $P_{IV}$}
We start with the symmetric $P_{IV}$ \cite{Noumi}:
\begin{equation}
\begin{split}
f_0'&=f_0 \left(f_1-f_2\right)+\alpha_0\, ,\\
f_1'&=f_1 \left(f_2-f_0\right)+\alpha_1\, ,\\
f_2'&=f_2\left(f_0-f_1\right)+\alpha_2\, ,
\label{symp4}
\end{split}
\end{equation}
where $f_i = f_i(z)$ and $'=d/dz$.

By summing these equations we get
\[f_0'+f_1'+f_2'=\alpha_0+\alpha_1+\alpha_2\]

Defining $\sigma:=\alpha_0+\alpha_1+\alpha_2$ and making one integration, we get
\begin{equation}\label{sumfp4}
f_0+f_1+f_2=\sigma z+\chi
\end{equation}
Setting the integration constant $\chi=0$ and eliminating $f_2=\sigma z-f_0-f_1$ from \eqref{symp4} we obtain:
\begin{equation}
\begin{split}\label{noumisum}
f_0'(z)=&f_0 \left(-\sigma z+f_0+2 f_1\right)+\alpha _0\,,\\
f_1'(z)=&f_1\left(\sigma z-2 f_0-f_1\right)+\alpha _1\,, 
\end{split}
\end{equation}
while the third equation in \eqref{symp4} can be obtained
by summing the above two equations.

Traditionally, either $\sigma$ is set to $1$ or it can be absorbed by the following transformations :
	\begin{equation}\label{absorb}
	\alpha_i=\tilde{\alpha_i}\sigma,\qquad f_i(z)=\sqrt{\sigma}\tilde{f_i}(\tilde{z}),\qquad z=\tilde{z}/\sqrt{\sigma}.
	\end{equation}

By further eliminating $f_1$ or $f_0$ from \eqref{noumisum}  we get
for the remaining component:
\begin{equation}\label{p4ff}
f_i''(z)=\frac{f_i'{}^2}{2 f_i}-\frac{\alpha _i^2}{2 f_i}
+\left(\frac{1}{2}\sigma^2 z^2+ {(-1)^i(2 \alpha _0+2 \alpha
	_1-\alpha_i- \sigma)}\right) f_i-2 \sigma  z f_i{}^2
+\frac{3}{2} f_i{}^3
,\quad \; i=0,1.
\end{equation}

\subsubsection{Degeneracies on the symmetric $P_{IV}$}
Here we formulate coalescence in the setting of the  symmetric P$_{IV}$ equations \eqref{symp4}
through the following transformations \cite{vccp4-2} :
\begin{equation}\label{te}
\begin{split}
f_i(z)&\to f_i(z)+\frac{1}{\epsilon},\qquad z\to z+\frac{2}{\sigma \epsilon^2},\\
\alpha_0&\to \epsilon \alpha_0 - \frac{1}{\epsilon^2},\qquad
\alpha_1\to \epsilon \alpha_1 + \frac{1}{\epsilon^2},\qquad
\alpha_2\to \epsilon \alpha_2 \, .
\end{split}
\end{equation}
Applying the above transformation to the first order equations \eqref{symp4} 
yields:
\begin{align}\label{p4coal}
f_0'(z)=&f_0 \left(f_1-f_2\right)+\frac{f_1-f_2}{\epsilon }+\epsilon \alpha_0 - \frac{1}{\epsilon^2}\nonumber\\
f_1'(z)=&f_1 \left(f_2-f_0\right)+\frac{f_2-f_0}{\epsilon }+\epsilon \alpha_1 + \frac{1}{\epsilon^2}\\
f_2'(z)=&f_2\left(f_0-f_1\right)+\frac{f_0-f_1}{\epsilon }+\epsilon \alpha_2\nonumber
\end{align}

Summing the equations above and making one integration, we arrive at
\begin{equation}\label{xi}
f_0+f_1+f_2=\epsilon\sigma z+\left(-\xi  \epsilon-\frac{1}{\epsilon }\right)
\end{equation}
that term in parenthesis is a suitably chosen integration constant corresponding to $\chi$ in \eqref{sumfp4}.

Eliminating $f_2$ we get:
\begin{equation}\label{mixed}
\begin{split}
f_0'(z)&=\epsilon  \left(\alpha _0-\sigma z f_0+\xi f_0\right)+\frac{2 f_0+2
	f_1}{\epsilon }+f_0^2+2 f_1 f_0-\sigma z+\xi\,,\\
f_1'(z)&=\epsilon  \left(\alpha _1+\sigma z f_1-\xi f_1\right)+\frac{-2 f_0-2
	f_1}{\epsilon }-f_1^2-2 f_0 f_1+\sigma z-\xi \, .
\end{split}
\end{equation}

By eliminating $f_1$ from \eqref{mixed} we obtain a second order ODE for $f_0(z)$ depending on $\sigma,\,\xi$ and $\epsilon$, which admits different solutions for different limits of these parameters.

\begin{itemize}
	\item Such an equation is the traditional $P_{IV}$ \eqref{p4ff} by absorbing $\xi$ to $z$, followed by \eqref{absorb} and absorbing $\epsilon$ from $f_0$ and $\alpha_i$ (inverse of \eqref{te});

\item if we take the limit $\sigma\to 0$ and absorbing $\epsilon$ from $f_0$ and $\alpha_i$ (inverse of \eqref{te}), we get the $I_{30}$ equation:
\begin{equation}\label{i30ff}
f_0''=\frac{f_0'{}^2}{2 f_0}+\frac{3}{2} f_0{}^3+f_0{}^2 \left(2 \xi  \epsilon -\frac{4}{\epsilon }\right)+f_0(z) \left(\alpha _0+2 \alpha _1-2 \xi +\frac{\xi ^2 \epsilon ^2}{2}+\frac{2}{\epsilon ^2}\right)-\frac{\alpha _0^2}{2f_0}
\end{equation}
\end{itemize}

Instead of the previous steps' limits, one keeps $\sigma$ and takes $\epsilon\to 0$ yielding:
\begin{equation}
f_0''(z)=2 f_0^3-2( \sigma  z-\xi) f_0 +2 \alpha_0+ 2\alpha_1-\sigma
\end{equation}
which is
\begin{itemize}
	\item the $P_{II}$ equation \eqref{p2} by either absorbing $\xi$ into $z$, or taking $\xi=0$;
	\item the $I_8$ equation \eqref{i8} by taking $\sigma=0$;
	\item the same $I_8$ if one had taken first the limit $\sigma\to0$ and then $\epsilon\to0$.
\end{itemize}

One can see that the use of the $\epsilon\sigma$ parameter here turns this procedure equivalent to \eqref{p4p2} by the relations \eqref{absorb}.

This whole procedure can be visualized as:
\begin{center}
	
\begin{tikzpicture}
\draw[very thick][<->]
(0,0) node[left] {$P_{IV}$}
--(2,0) node[pos=0.5,above] {($\epsilon,\sigma,\xi$)}
node[right] {$P_{IV}(\epsilon,\sigma,\xi)$};
\draw[very thick][->]
(4.2,0) -- (6.0,0) node[pos=0.5,above] {$\epsilon\to0$}
node[right] {$P_{II}(\sigma,\xi)$};
\draw[thick][->]
(3.1,-0.3) -- (3.1,-2) node[above,pos=0.5, rotate=270] {$\sigma\to0$}
node[below]{$I_{30}(\epsilon,\xi)$};
\draw[very thick][<->]
(0,-2.3) node[left]{$I_{30}(\epsilon,\xi)$}
 -- (2,-2.3) node[pos=0.5,above] {$\overleftrightarrow{\epsilon}$};
\draw[thick][->]
(6.8,-0.3) -- (6.8,-2) node[above,pos=0.5, rotate=270] {$\sigma\to0$}
node[below]{$I_{8}$};
\draw[very thick][->]
(4.2,-2.3) -- (6.3,-2.3) node[pos=0.5,above] {$\epsilon\to0$};
\end{tikzpicture}
\end{center}

where $P_{IV}$ is \eqref{p4ff}; ($\epsilon,\sigma,\xi$) is \eqref{te} with \eqref{xi}; $P_{IV}(\epsilon,\sigma,\xi)$ is the second order ODE from \eqref{mixed}; $\overleftrightarrow{\epsilon}$ is \eqref{te} without the $z$ transformation; and the leftmost $I_{30}(\epsilon,\xi)$ is the equation \eqref{i30ff}.
\subsection{$P_V$}
\subsubsection{The symmetric $P_V$}
Here the traditional symmetric $P_V$ is described as in the literature and also the coalescence to $I_{38}$ is presented.

The symmetric equations for $P_V$, as described by Noumi \cite{Noumi}, correspond to the system  of differential equations:

\begin{equation}
\begin{split}\label{p5sym}
z f_0'=\left(\frac{\sigma }{2}-\alpha _2\right) f_0+\alpha _0 f_2+\left(f_1- f_3\right) f_2f_0\\
z f_1'=\left(\frac{\sigma }{2}-\alpha _3\right) f_1+\alpha _1 f_3+\left(f_2-f_0\right)f_3 f_1\\
z f_2'=\left(\frac{\sigma }{2}-\alpha _0\right) f_2+\alpha _2 f_0+\left(f_3- f_1\right)f_0f_2\\
z f_3'=\left(\frac{\sigma }{2}-\alpha _1\right) f_3+\alpha _3 f_1+\left(f_0- f_2\right)f_1 f_3
\end{split}
\end{equation}
where $\sigma:=\alpha_0+\alpha_1+\alpha_2+\alpha_3$.

The following change of variables is equivalent to rescale $\sigma\to 1$:
\begin{equation}
f_i(z)= \sqrt{\sigma}\tilde{f}_i(x), \qquad z= (\sigma x)^{1/\sigma},\qquad\alpha_i=\sigma\tilde{\alpha}_i, \qquad i=0,1,2,3.
\end{equation}

By summing the first and third (resp. second and fourth) equations in \eqref{p5sym} we get, respectively:
\begin{gather}
z f_0'+zf_2'=\frac{\sigma}{2}(f_0+f_2)\nonumber\\
z f_1'+zf_3'=\frac{\sigma}{2}(f_1+f_3)\label{sumfa3}
\end{gather}
we are able to perform 2 integrations, therefore obtaining 2 integration constants, $\epsilon_0$ and $\epsilon_1$:
\begin{gather}
f_0+f_2=\epsilon_0z^{\sigma/2}(f_0+f_2)\\
f_1+f_3=\epsilon_1z^{\sigma/2}(f_1+f_3)
\end{gather}
$\epsilon_0$ and $\epsilon_1$ are normally set to 1, but for reasons explained ahead, they will be kept here.

By eliminating $f_2$ and $f_3$ in \eqref{p5sym} using the above relations, we get for $f_0$:
\begin{align}
z f_0'=&\alpha _0 \left(\epsilon_0 z^{\sigma/2}-f_0\right)+f_0 \left(f_1 \left(\epsilon_0 z^{\sigma/2}-f_0\right)-\left(\epsilon_0 z^{\sigma/2}-f_0\right) \left(\epsilon_1 z^{\sigma/2}-f_1\right)\right)+f_0 \left(\frac{\sigma}{2}-\alpha _2\right)\\
z f_1'=&\alpha _1 \left(\epsilon_1 z^{\sigma/2}-f_1\right)+f_1 \left(\left(\epsilon_0 z^{\sigma/2}-f_0\right) \left(\epsilon_1 z^{\sigma/2}-f_1\right)-f_0 \left(\epsilon_1 z^{\sigma/2}-f_1\right)\right)+f_1 \left(\frac{\sigma}{2}-\alpha _3\right)
\end{align}
we are able to solve the above equations either for $f_0$ or for $f_1$, yielding second order differential equations. For simplicity, only $f_0$ will be shown hereinafter, and it gets the form:
\begin{equation}
f_0''=\frac{f_0'{}^2}{2}\left(\frac{1}{f_0}-\frac{1}{\left(\epsilon_0 z^{\sigma/2}-f_0\right)}\right)-\frac{f_0'}{z}+...\label{f0cru}\\
%f_1''=\frac{f_1'{}^2}{2}\left(\frac{1}{f_1}-\frac{1}{\left(\epsilon_1 z^{\sigma/2}-f_1\right)}\right)-\frac{f_1'}{z}+...\label{f1cru}
\end{equation}
such form of equation suggests it is a Painleve equation, and we should perform the following change of variables, as described in Ince's book:
\begin{equation}\label{gi}
f_0(z)=\frac{\epsilon_0 z^{\sigma/2}}{1-g_0(z)},%\qquad f_1(z)=\frac{\epsilon_1 z^{\sigma/2}}{1-g_1(z)}.
\end{equation}
and in order to eliminate some powers of $\sigma$, we also perform the transformation:
\begin{equation}\label{sigmatransf}
z=x^{1/\sigma}.
\end{equation}
Such change of variables reveals us it is a $P_V$ equation with parameters, respectively:
\begin{equation}
%\begin{split}
g_0:\qquad a = \frac{\alpha _0^2}{2 \sigma ^2},\quad b = -\frac{\alpha _2^2}{2 \sigma ^2},\quad c = \frac{\left(\alpha _3-\alpha _1\right) \epsilon_0 \epsilon_1}{\sigma ^2},\quad d = -\frac{\epsilon_0^2 \epsilon_1^2}{2 \sigma ^2}%\\
%g_1:\qquad a = \frac{\alpha _1^2}{2 \sigma ^2},\quad b = -\frac{\alpha _3^2}{2 \sigma ^2},\quad c = \frac{\left(\alpha _0-\alpha _2\right) \epsilon_0 \epsilon_1}{\sigma ^2},\quad d = -\frac{\epsilon_0^2 \epsilon_1^2}{2 \sigma ^2}
%\end{split}
\end{equation}
one can notice that the parameters above are related by an index shift.
\\

The case for $\sigma=0$ is also immediately noticeable, since it appears as a singularity in step \eqref{sigmatransf}.

In such case we have to go back to the $g_0$ equation \eqref{gi} still in the $z$ variable and then set $\sigma\to0$ followed by the transformation
\[z=e^x\]
This yields the $I_{38}$ equation \eqref{i38} with parameters, respectively:
\begin{gather}
g_0:\qquad\alpha _0= -\sqrt{2} \sqrt{a},\quad\alpha _1= \frac{1}{4} \left(2 \sqrt{2} \left(\sqrt{a}+i \sqrt{b}\right)-\frac{2 c}{\epsilon_0 \epsilon_1}-\epsilon_0 \epsilon_1\right),\quad\alpha _2= -i \sqrt{2} \sqrt{b},\nonumber
\\\alpha _3= \frac{1}{4} \left(2 \sqrt{2} \left(\sqrt{a}+i \sqrt{b}\right)+\frac{2 c}{\epsilon_0 \epsilon_1}+\epsilon_0 \epsilon_1\right),\quad d= -\epsilon_0^2 \epsilon_1^2%\\
%g_1:\qquad\alpha _0= \frac{1}{4} \left(2 \sqrt{2} \left(\sqrt{a}+i \sqrt{b}\right)+\frac{2 c}{\epsilon_0 \epsilon_1}+\epsilon_0 \epsilon_1\right),\quad\alpha _1= -\sqrt{2} \sqrt{a},\nonumber\\
%\alpha _2= \frac{1}{4} \left(2 \sqrt{2} \left(\sqrt{a}+i \sqrt{b}\right)-\frac{2 c}{\epsilon_0 \epsilon_1}-\epsilon_0 \epsilon_1\right),\quad \alpha _3= -i \sqrt{2} \sqrt{b},\quad d= -\epsilon_0^2 \epsilon_1^2.
\end{gather}

\subsubsection{Coalescence to $P_{III}$}
An effective and very symmetrical way of obtaining the coalescence from $P_V$ to $P_{III}$ is by adding terms $\beta_0(f_1+f_3)$ and $\beta_1(f_0+f_2)$, such that:
\begin{equation}
\begin{split}\label{p5symdef}
z f_0'=\left(\frac{\sigma }{2}-\alpha _2\right) f_0+\alpha _0 f_2+\left(f_1- f_3\right) f_2f_0-\beta_0(f_1+f_3)\\
z f_1'=\left(\frac{\sigma }{2}-\alpha _3\right) f_1+\alpha _1 f_3+\left(f_2-f_0\right)f_3 f_1+\beta_1(f_0+f_2)\\
z f_2'=\left(\frac{\sigma }{2}-\alpha _0\right) f_2+\alpha _2 f_0+\left(f_3- f_1\right)f_0f_2+\beta_0(f_1+f_3)\\
z f_3'=\left(\frac{\sigma }{2}-\alpha _1\right) f_3+\alpha _3 f_1+\left(f_0- f_2\right)f_1 f_3-\beta_1(f_0+f_2)
\end{split}
\end{equation}
such approach was first developed in \cite{p3-5}.
\\

We repeat the same steps as in the traditional case up to \eqref{sigmatransf}; namely, eliminating $f_2$ and $f_3$ from the system, solving it for $f_0''$ and $f_1''$, making the variable change \eqref{gi} followed by \eqref{sigmatransf}, we end up with a $P_V$ equation with parameters:
\begin{equation}
%\begin{split}
g_0:\qquad a = \frac{\left(\alpha _0 \epsilon_0+\epsilon_1 \beta_0\right){}^2}{2 \sigma ^2},\quad b = -\frac{\left(\alpha _2 \epsilon_0-\epsilon_1 \beta_0\right){}^2}{2 \sigma ^2},\quad c = \frac{\epsilon_0 \left(\left(\alpha _3-\alpha _1\right) \epsilon_1+2 \epsilon_0 \beta_1\right)}{\sigma ^2},\quad d = -\frac{\epsilon_0^2 \epsilon_1^2}{2 \sigma ^2}%\\
%g_1:\qquad a = \frac{\left(\alpha _1 \epsilon_1-\epsilon_0 \beta_1\right){}^2}{2 \sigma ^2},\quad b = -\frac{\left(\alpha _3 \epsilon_1+\epsilon_0 \beta_1\right){}^2}{2 \sigma ^2},\quad c = \frac{\epsilon_1 \left(\left(\alpha _0-\alpha _2\right) \epsilon_0+2 \epsilon_1 \beta_0\right)}{\sigma ^2},\quad d = -\frac{\epsilon_0^2 \epsilon_1^2}{2 \sigma ^2}
%\end{split}
\end{equation}

That is, the addition of $\epsilon_i$ terms is equivalent to:
\[\alpha_0\to \alpha_0\epsilon_0+\beta_0\epsilon_1,\quad
\alpha_1\to \alpha_1\epsilon_1-\beta_1\epsilon_0,\quad
\alpha_2\to \alpha_2\epsilon_0-\beta_0\epsilon_1,\quad
\alpha_3\to \alpha_3\epsilon_1+\beta_1\epsilon_0.\]

Now, if we instead of making the transformations \eqref{gi} when we arrive at the second-order equation with the form \eqref{f0cru}, we take $\epsilon_0\to0$ (respectively $\epsilon_1\to0$ for the $f_1$ equation), we arrive at the equation:
\begin{equation}
f_0''=\frac{f_0'{}^2}{f_0}-\frac{f_0'}{z}+\epsilon_1^2 z^{\sigma -2} \left(f_0{}^3-\frac{\beta_0^2}{f_0}\right)+\epsilon_1 z^{\frac{\sigma }{2}-2} \left(\beta_0 \left(-\alpha _1-\alpha _3+\sigma \right)+\left(\alpha _3-\alpha _1\right) f_0(z){}^2\right)\label{f0r0e}%\\
%f_1''=&\frac{f_1'{}^2}{f_1}-\frac{f_1'}{z}+\epsilon_0^2 z^{\sigma -2} \left(f_1{}^3-\frac{\beta_1^2}{f_1}\right)-\epsilon_0 z^{\frac{\sigma }{2}-2}
%\left(\left(\alpha _1+\alpha _3\right) \beta_1+f_1^2 \left(\alpha _1+2 \alpha _2+\alpha _3-\sigma \right)\right) \label{f1r1e}
\end{equation}
that can have the powers of $\sigma$ eliminated by the transformation
\[z=x^{2/\sigma}\]
yielding the traditional $P_{III}$ equation \eqref{p3} with parameters:
\begin{equation}
%\begin{split}
f_0:\quad a=\frac{4 \left(\alpha _3-\alpha _1\right) \epsilon_1}{\sigma ^2},\quad b = \frac{4 \epsilon_1 \beta_0 \left(\alpha_0+\alpha _2 \right)}{\sigma ^2},\quad c= \frac{4 \epsilon_1^2}{\sigma ^2},\quad d= -\frac{4 \epsilon_1^2 \beta_0^2}{\sigma ^2},%\\
%f_1:\qquad a=\frac{4 \epsilon_0 \left(\alpha_0-\alpha_2 \right)}{\sigma ^2},\quad b= -\frac{4 \left(\alpha _1+\alpha _3\right) \epsilon_0 \beta_1}{\sigma ^2},\quad c= \frac{4 \epsilon_0^2}{\sigma ^2},\quad d= -\frac{4 \epsilon_0^2 \beta_1^2}{\sigma ^2}
%\end{split}\label{symp3}
\end{equation}
as usual, we notice that the point $\sigma=0$ is a singularity, therefore setting $\sigma=0$ on \eqref{f0r0e}% and \eqref{f1r1e}
, followed the variable change $z=e^x$, it becomes $I_{12}$ \eqref{i12} with parameters:

\begin{equation}
%\begin{split}
f_0:\qquad a= \epsilon_1^2,\quad b=\epsilon_1 \left(\alpha _3-\alpha _1\right),\quad c= -\epsilon_1\beta_0\left(\alpha _1+\alpha _3\right),\quad d= -\epsilon_1^2 \beta_0^2%\\
%f_1:\qquad a= \epsilon_0^2,\quad b= \epsilon_0\left(\alpha _0-\alpha _2\right) ,\quad c= -\epsilon_0\beta_1\left(\alpha _1+\alpha _3\right),\quad d= -\epsilon_0^2 \beta_1^2
%\end{split}\label{symi12}
\end{equation}

The complementary case, of $f_{0,zz}$ with $\epsilon_1=0$ %and $f_{1,zz}$ with $\epsilon_0=0$ are 
is trivial since this limit is not a singularity, therefore it is just $P_V$ (or $I_{38}$ when $\sigma=0$) equation.

%One can also immediately see from \eqref{symp3} and \eqref{symi12} that the case with $\beta_0=\beta_1=0$ give us the incomplete equations. One can also prove that these 2 equations, $P_{III}[b=d=0]$ and $I_{12}[c=d=0]$, are equivalent.

If we translate that process of going from $P_V$ to $P_{III}$ in literature's language, that is, calling $\epsilon_0=\epsilon$ and making the appropriate relabeling of the other parameters, this process is exactly like the old-fashioned coalescence limit \eqref{p5p3}.

Equations $I_{38}(g_0(x))$ and $I_{12}(f_0(x))$ are also connected by the transformations already described.

This whole procedure can be visualized as:
	
\begin{center}
	\begin{tikzpicture}
%\draw[very thick][->]
%(0,0) node[left] {$P_{V}(\sigma)$}
%--(2,0) node[pos=0.5,above] {$r_i\to0$}
%node[right] {$I_{12}[c=d=0]$};
%\draw[thick][->]
%(-0.7,-0.3) -- (-0.7,-2) node[above,pos=0.5, rotate=270] {$\sigma\to0$}
%node[below]{$I_{38}$};
\draw[very thick][<->]
(4,0) node[left]{$P_V(g_0(z);\sigma,\beta_i)$}
--(6,0)  node[pos=0.5,above] {$f_0\leftrightarrow g_0$};
\draw[very thick][<->]
(4,-2.3)--(6,-2.3) node[pos=0.5,above] {$f_0\leftrightarrow g_0$};
\draw[thick][->]
(2.6,-0.3) -- (2.6,-2) node[above,pos=0.5, rotate=270] {$\sigma\to0$}
node[below]{$I_{38}(g_0(z))$};
\draw[thick][<->]
(2.6,-2.8) -- (2.6,-4.5) node[above,pos=0.5, rotate=270] {$z=e^x$}
node[below]{$I_{38}(g_0(x))$};
\draw[very thick][->]
(9,0) node[left] {$P_{V}(f_0(z);\sigma,\beta_i)$}
--(11,0) node[pos=0.5,above] {$\epsilon_0\to0$}
node[right] {$P_{III}(f_0(z))$};
\draw[very thick][->]
(9,-2.3)--(11,-2.3) node[pos=0.5,above] {$\epsilon_0\to0$};
\draw[thick][->]
(7.6,-0.3) -- (7.6,-2) node[above,pos=0.5, rotate=270] {$\sigma\to0$}
node[below]{$I_{38}(f_0(z))$};
\draw[thick][->]
(12.2,-0.3) -- (12.2,-2) node[above,pos=0.5, rotate=270] {$\sigma\to0$}
node[below]{$I_{12}(f_0(z))$};
\draw[thick][<->]
(12.2,-2.8) -- (12.2,-4.5) node[above,pos=0.5, rotate=270] {$z=e^x$}
node[below]{$I_{12}(f_0(x))$};
\end{tikzpicture}
\end{center}

where $P_V(f_0(z);\sigma,\beta_i)$ is the equation with the form \eqref{f0cru} originated from the system \eqref{p5symdef}; $f_0\leftrightarrow g_0$ is the transformation \eqref{gi}; and since specifically $I_{38}(g_0(x))$ and $I_{12}(f_0(x))$ that have the form described in literature they feature here too. 

\subsection{The B\"acklund Transformations for $\sigma=0$}
The theory for the behavior of the affine Weyl groups of the B\"acklund Transformations for the Painlev\'e equations under coalescence is well described in \cite{takano} and can be seen to some extent here too by noticing how the parameters combine to become the resulting ones eliminating singularities.

The result of taking $\sigma\to 0$ on B\"acklund transformations is different from what happens in coalescence, there the group structure shrinks and degenerates into a subgroup at each step of coalescence; here the group structure is still present, but totally spoiled due to an extra relation that does not allow the generations of an infinite chain of solutions, as will be seen.

Since the conclusion here is simple we take as a case study the symmetric $P_{IV}$, which is invariant by the B\"acklund transformations \cite{Noumi}:
\begin{equation}
\begin{array}{c|ccc|ccc}
{} & {\alpha_0} & { \alpha_1} &
{ \alpha_2} & { f_0} & { f_1} &
{ f_2}\\
\hline
{ s_0}&{ -\alpha_0} &{
	\alpha_1+\alpha_0} &{ \alpha_2+\alpha_0} &{
	f_0} &{ f_1+\frac{\alpha_0}{f_0}} &{
	f_2-\frac{\alpha_0}{f_0}}\\[2mm]
\hline
{ s_1}&{ \alpha_0+\alpha_1} &{
	-\alpha_1} &{ \alpha_2+\alpha_1} &{
	f_0-\frac{\alpha_1}{f_1}} &{ f_1} &{
	f_2+\frac{\alpha_1}{f_1}} \\[2mm]
\hline
{ s_2}&{ \alpha_0+\alpha_2} &{
	\alpha_1+\alpha_2} &{-\alpha_2} &{
	f_0+\frac{\alpha_2}{f_2}} &{ f_1-\frac{\alpha_2}{f_2}} &{
	f_2}
\end{array}
\label{p4bt}
\end{equation}
\[(s_i)^2=1,\qquad s_is_js_i=s_js_is_j\]

The above table should be read as $s_0(\alpha_0)=-\alpha_0$ , $s_0(\alpha_1)=\alpha_1+\alpha_2$ and so on, meaning that applying $s_i$ keeps system \eqref{symp4} invariant.

This implies that if one has a solution (obtained by any means) for $f_i(z)$ with constants $\{\alpha_0,\alpha_1,\alpha_2\}$ for $P_{IV}$ \eqref{p4ff}, we are immediately able to obtain a new solution with the new set of constants, and by combining the transformations $s_i$, it is easy to see that every time the constants combine into $\alpha_0+\alpha_1+\alpha_2=\sigma\neq 0$ such procedure can be continued indefinitely always summing $\sigma$ after some steps, creating an infinite chain of solutions for $P_{IV}$. For more details the author refers to \cite{Noumi}.

One notices that the $\sigma$-parameter plays no evident role here, so using
\[\alpha_2=-\alpha_0-\alpha_1\] 
%which corresponds to \eqref{te} without the $\epsilon$ and with $\sigma=0$ 
and after plugging it in the transformations above, they become:
%\begin{equation}
%\begin{array}{c|cc|cc|}
%	{} & {\alpha_0} & { \alpha_1} &
%	{ f_0} & { f_1}\\
%	\hline
%	{ s_0}&{ -\alpha_0} &{
%		\alpha_1+\alpha_0}  &{
%		f_0} &{ f_1+\frac{\alpha_0}{f_0}}\\[2mm]
%	\hline
%	{ s_1}&{ \alpha_0+\alpha_1} &{
%		-\alpha_1} &{
%		f_0-\frac{\alpha_1}{f_1}} &{ f_1}\\[2mm]
%	\hline
%	{ s_2}&{ -\alpha_1} &{
%		-\alpha_0} &{
%		f_0+\frac{\alpha_2}{f_2}} &{ f_1-\frac{\alpha_2}{f_2}}\\[2mm]
%	\hline
%	%{ \pi}&{ \alpha_1} &{
%	%	\alpha_2} &{ \alpha_0} &{
%	%	f_1} &{ f_2} &{ f_0} 
%\end{array}
%\label{i30bt}
%\end{equation}
\begin{equation}
\begin{array}{c|cc}
{} & {\alpha_0} & { \alpha_1} \\
\hline
{ s_0}&{ -\alpha_0} &{
	\alpha_1+\alpha_0} \\[2mm]
\hline
{ s_1}&{ \alpha_0+\alpha_1} &{
	-\alpha_1}\\[2mm]
\hline
{ s_2}&{ -\alpha_1} &{
	-\alpha_0}
\end{array}
\label{i30bt}
\end{equation}
one can easily see that the transformations now are not independent anymore:
\begin{equation}\label{spoil}
s_is_js_i=s_k,\qquad(\text{or}\quad s_is_j=s_ks_i)\qquad i\neq j\neq k
\end{equation}
since relation $(s_i)^2=1$ and $s_is_js_i=s_js_is_j$ are still valid, we see that the other possibility of combination of $s_i$ also degenerates to this simpler case:
\begin{equation}
s_is_js_k=s_ks_is_k=s_j
\end{equation}
therefore all possible combinations of constants are restricted to $\pm\alpha_i$ and $\pm\alpha_i\pm\alpha_j$.

Since one cannot generate a chain of solutions, one neither can create a chain of rational solutions nor associate a polynomial for them, like the Hermite polynomials that would be associated with the rational solutions of $P_{IV}$ \cite{gromak}, in this example. Since it applies to $I_{12}$ (and to the others autonomous equations), this is the of reason such a chain of rational solutions does not appear in Jacobi Elliptic functions.

The generalization of \eqref{spoil} to higher $N$ can be written as:
\begin{equation}
(s_{i+1}\ldots s_{i+N-1})s_{i+N}=s_{i}(s_{i+1}\ldots s_{i+N-1})
\end{equation}
and similar observations apply.
\section{Conclusion}
Hybrid Painlev\'e equations have been a goal of several researchers, e.g. 
\cite{kudryashov,rogers}, and to expand equations with their possible degeneracy parameters allows one to obtain a hybrid equation with both a Painlev\'e equation and an elliptic equation, for example. This was done in \cite{p3-5},\cite{vcclimits} and \cite{vccp4-2} and here the complete framework for such equations was provided since the basic recipe there was to first find the coalescence in the framework of symmetric equations and then extend the parameter space to all possible constants of integration the system provides.

The coalescence cascades were seen here to be preserved for multiple properties and reductions, like for autonomous equations, Riccati equations, symmetric equations, and B\"acklund transformations.

The limit to autonomous equations here explored ($\sigma\to0$) is a bridge from Painlev\'e equations to any simpler ODE by this systematic procedure of coalescence, and can be applied to a multitude of properties of these equations, even possibly for numerical algorithms.

\section{Acknowledgements}

The author acknowledges the research support of the S\~ao Paulo Research Foundation (FAPESP) via grant number 2016/22122-9 and 2019/03092-0.

\end{document}